\documentclass[twocolumn]{aastex631}
\usepackage{amsmath,amssymb,mhchem}
\shortauthors{Glidden et al.}
\shorttitle{Isotopologues}

\begin{document}

\title{Can Isotopologues Be Used as Biosignature Gases in Exoplanet Atmospheres?}

\correspondingauthor{Ana Glidden}
\email{aglidden@mit.edu}

\author[0000-0002-5322-2315]{Ana Glidden}
\affiliation{Department of Earth, Atmospheric and Planetary Sciences, Massachusetts Institute of Technology, Cambridge, MA 02139, USA}
\affiliation{Kavli Institute for Astrophysics and Space Research, Massachusetts Institute of Technology, Cambridge, MA 02139, USA}

\author[0000-0002-6892-6948]{Sara Seager}
\affiliation{Department of Earth, Atmospheric and Planetary Sciences, Massachusetts Institute of Technology, Cambridge, MA 02139, USA}
\affiliation{Department of Physics and Kavli Institute for Astrophysics and Space Research, Massachusetts Institute of Technology, Cambridge, MA 02139, USA}
\affiliation{Department of Aeronautics and Astronautics, Massachusetts Institute of Technology, Cambridge, MA 02139, USA}

\author[0000-0002-1921-4848]{Janusz J. Petkowski}
\affiliation{Department of Earth, Atmospheric and Planetary Sciences, Massachusetts Institute of Technology, Cambridge, MA 02139, USA}
\affiliation{JJ Scientific, Mazowieckie, 02-792 Warsaw, Poland}
\affiliation{Faculty of Environmental Engineering, Wroclaw University of Science and Technology, 50-370 Wroclaw, Poland}

\author[0000-0002-1348-9584]{Shuhei Ono}
\affiliation{Department of Earth, Atmospheric and Planetary Sciences, Massachusetts Institute of Technology, Cambridge, MA 02139, USA}

\begin{abstract}
Isotopologue ratios are anticipated to be one of the most promising signs of life that can be observed remotely. On Earth, carbon isotopes have been used for decades as evidence of modern and early metabolic processes. In fact, carbon isotopes may be the oldest evidence for life on Earth, though there are alternative geological processes that can lead to the same magnitude of fractionation. However, using isotopologues as biosignature gases in exoplanet atmospheres presents several challenges. Most significantly, we will only have limited knowledge of the underlying abiotic carbon reservoir of an exoplanet. Atmospheric carbon isotope ratios will thus have to be compared against the local interstellar medium or, better yet, their host star. A further substantial complication is the limited precision of remote atmospheric measurements using spectroscopy. The various metabolic processes which cause isotope fractionation cause less fractionation than anticipated measurement precision (biological fractionation is typically 2 to 7\%). While this level of precision is easily reachable in the laboratory or with special \textit{in situ} instruments, it is out of reach of current telescope technology to measure isotope ratios for terrestrial exoplanet atmospheres. Thus, gas isotopologues are poor biosignatures for exoplanets given our current and foreseeable technological limitations.
\end{abstract}

\section{Introduction} \label{intro}

As a human species, we have long gazed at the heavens and wondered if we are alone. Despite the thousands of exoplanets discovered over the last decade, we still have not found any definitive signs of life on other planets. However, the technology gap limiting detections of exoplanet atmospheric biosignature gases may be closing with the recent launch of JWST \citep{Gardner2006}. For the first time, we can now measure the atmospheric constituents and properties of potentially habitable, rocky worlds. In anticipation of our upcoming observations, we must sort out the pros and cons of possible biosignature gases.

Carbon isotope data are potentially the oldest geochemical evidence for ancient life on Earth \citep{Mojzsis1996, Eiler1997}. Biotic deposits of carbon have a higher ratio of $^{12}$C-to-$^{13}$C than abiotic deposits \citep{Rothschild1989}. Differences in carbon isotope ratios between inorganic and organic carbon sources as far back as 3.5 Ga could be evidence of a metabolic processes like photosynthesis, though this is not conclusive \citep[][and references therein]{Garcia2021}. \citet{Bell2015} measured isotopically light graphite preserved in a 4.1 Ga zircon $\delta^{13}$C$_{\textrm{PDB}}=$ $-24\pm5$\textperthousand\, where PDB refers to the Pee Dee Belemnite standard terrestrial reference carbon isotope ratio. The measurement could be indicative of early enzymatic carbon fixation, though there are plausible abiotic false positives such as post-depositional processes like metamorphism \citep{Bell2015, Eiler1997}. Intriguingly, over this vast timescale, the difference between the abiotic and biotic reservoirs has remained largely constant despite the evolution of life on Earth, variations in rainfall, and changes in the partial pressure of atmospheric carbon dioxide \citep{Garcia2021,Hare2018}. Today, we can measure small seasonal changes in the isotopic composition of key photosynthetic gases \citep{Keeling1958}. Given our success using isotopic compositions as evidence of past and current life on Earth, isotopic data are also thought to be among the strongest signs of life that can be remotely detected on other planets \citep{Neveu2018}.

Before we can consider using isotopes as a sign of life on other planets, we must evaluate if it will be possible to detect isotopologues at all---whether their fractionation is a byproduct of biological or planetary processes. As such, we will focus this review mainly on the detectability of isotopic spectral signatures that can be created through biological processes rather than discussing those processes in detail. Isotopologues are molecules that contain one or more isotope substitutions. While the number of protons determines which element is which, the same element can have different numbers of neutrons. Elements with different numbers of neutrons are referred to as isotopes. Molecules made of the same elements, but with different isotopes of those elements are called isotopologues. For example, $^{1}$H$_{2}^{16}$O and $^{1}$H$^{2}$H$^{16}$O (also known as HDO) are both isotopologues of water. The successful detection of an isotopologue gas in an exoplanet atmosphere requires a large aperture, high spectral resolution telescope capable of observing the spectral separation between isotopologues. In the last few years, simulations have assessed the possibility of detecting isotope fractionation caused by planetary processes using near-future instruments on JWST, Very Large Telescope (VLT), and Extremely Large Telescope (ELT) \citep{Morley2019, Lincowski2019, MolliereSnellen2019}. Simulations showed the deuterium-to-hydrogen ratio was detectable for cool brown dwarfs with JWST with CH$_{3}$D, but not with HDO \citep{Morley2019}. Furthermore, simulations showed that $^{18}$O/$^{16}$O could be measured in the atmospheres of the terrestrial planets TRAPPIST-1~b and d with upcoming JWST transmission observations and could be used as evidence of ocean loss \citep{Lincowski2019}. \citet{MolliereSnellen2019} found that $^{13}$CO was detectable in the atmospheres of hot Jupiters from the ground with the CRIRES+ instrument on the VLT, while HDO and CH$_{3}$D were accessible with METIS on the ELT for self-luminous planets. In 2021, the pioneering work of \citet{Zhang2021} successfully detected the isotopologue $^{13}$CO in the atmosphere of a young super-Jupiter using ESO's Very Large Telescope. Soon after, \citet{Line2021} detected $^{13}$CO in the atmosphere of WASP-77 A b using the Gemini South Observatory. While $^{13}$CO has been successfully detected in two exoplanet atmospheres using ground-based instruments, $^{13}$CO$_{2}$ is an objectively better target for space-based observations with JWST and will play an important roll in observational programs of future high resolution giant ground-based telescopes.

The ratio $^{13}$CO$_{2}$/$^{12}$CO$_{2}$ presents the best opportunity to detect evidence for carbon isotope fractionation in an exoplanet's atmosphere, including biological fractionation, with JWST. The strong CO$_{2}$ feature around 4.3 $\mu$m has only recently become accessible due to JWST \citep{Ahrer2023}. In anticipation of JWST observations, we evaluated the detectability of $^{13}$CO$_{2}$ in the atmospheres of temperate sub-Neptunes as evidence of an aerial biosphere \citep{Glidden2022}. We were motivated to explore measuring $^{13}$CO$_{2}$ as a biosignature gas as CO$_{2}$ is an incredibly strong absorber, making it easier to detect its presence than most other molecules. Additionally, no other metabolically-generated gas has as large of a spectral separation between the maximum intensity of its first and second most abundant isotopologue as CO$_{2}$ with a maximum separation of 0.121$\mu$m compared with 0.105$\mu$m for CO and 0.010$\mu$m for CH$_{4}$ within 1 to 5$\mu$m. Furthermore, for temperate planets, CO$_{2}$ will be more abundant than CO and CH$_{4}$ if the planet's atmosphere is in chemical equilibrium. Thus, our best chance at detecting metabolically-driven isotope fractionation is via $^{13}$CO$_{2}$. Unfortunately, we found that $^{13}$CO$_{2}$ could only be distinguished from $^{12}$CO$_{2}$ using JWST for the most idealized case of a sub-Neptune or larger sized planet with a large scale height atmosphere around a small, bright star \citep{Glidden2022}. 

Here, we further evaluate the usefulness of isotopologues---in particular $^{13}$CO$_{2}$---as biosignature gases in exoplanetary atmospheres in general. In Section \ref{solarsystem}, we consider how isotopes may provide evidence of metabolic processes in the solar system. Next, in Section \ref{exoplanets}, we discuss the current status of detecting isotopologues in exoplanet atmospheres. Then, in Section \ref{challenges}, we evaluate the difficulties associated with using isotopologues as biosignature gases. In particular, we discuss the importance of establishing a baseline value for isotope isotope ratios; abiotic and biotic isotope fractionation processes and how they can be distinguished; and the limited number of suitable targets. In Section \ref{solutions}, we outline possible solutions for using isotopologues as supportive evidence for life in exoplanet atmospheres. Finally, in Section \ref{conclusion}, we summarize our findings.

\section{Isotopes as Biosignatures in the Solar System: Possible Evidence of Metabolic Processes on Nearby Worlds from Surface and Atmospheric Measurements} \label{solarsystem}

Isotopologues have long been considered possible bioindicators on solar system bodies. Over three decades ago, \citet{Rothschild1989} investigated using carbon isotope ratios as evidence of life on Mars. Understanding how isotopologues have been used as bioindicators in our solar system provides the foundation for their potential future use as biosignature gases in exoplanet atmospheres. However, most simulations and measurements of isotopes on Mars have focused on abiotic processes.

Carbon isotope ratios for Mars have been measured many times using observations from Earth, gas trapped in Martian meteorites, and \textit{in situ} by Martian rovers \citep[e.g.,][]{Owen1982, Schrey1986, Niles2010, Webster2013, Mahaffy2013, Leshin2013}. Atmospheric measurements taken by Mars landers of atmospheric CO$_{2}$ have been used to interpret geological processes \citep[e.g.,][]{Niles2010, Webster2013, Mahaffy2013}. Here, we will discuss some of the most recent measurements.

Recently, \citet{Alday2021} used solar occultations to measure the Martian atmospheric isotopic composition of CO$_{2}$ to better understand the history of Mars' climate. They found largely Earth-like isotope compositions for both oxygen and carbon, and they focus their discussion on how their measurements inform on the atmospheric escape of carbon and photochemical processes rather than positing any potential metabolic explanation. \citet{Alday2021} suggest that $20\%$--$40\%$ of Mars' atmospheric carbon has been lost to space. 

Several processes are responsible for the escape of carbon to space, including photodissociation of CO, dissociative recombination of CO+, and sputtering by pick-up ions \citep[e.g.,][and references therein]{Alday2021, Jakosky2019, Hu2015}. Processes such as photodissociation, dissociative recombination, and sputtering also  preferentially remove the lighter $^{12}$C, enriching Mars' atmosphere in $^{13}$C over time \citep{Jakosky2019}. Modeling by \citet{Hu2015} predicted the impact of escape processes on atmospheric isotopic composition. Additionally, differences in the cross-sections of CO$_{2}$ isotopologues can lead to photoinduced isotope fractionation \citep{Schmidt2013}. 

Using the Sample Analysis at Mars instrument on the Curiosity Rover, \citet{House2022} measured a large range of carbon isotope ratios using methane gas released from surface samples from $\delta^{13}$C$_{\textrm{PDB}}=$ $-137\pm8$\textperthousand\ to $22\pm10$\textperthousand. Of their measurements, ten had a $\delta\ce{^{13}C}$ of less than $-$70\textperthousand. If such a depletion in $^{13}$C was measured on Earth, it would be seen as evidence of a past microbial metabolic process. \citet{House2022} attribute the paucity of $^{13}$C in the methane samples analyzed to three possible mechanisms: (1) microbial oxidation of methane, (2) deposited interstellar dust from passing through a giant molecular cloud depleted in $^{13}$C, or (3) abiotic photochemical reduction of CO$_{2}$. The surface measurements made by \citet{House2022} were unable to conclusively prove nor rule out the presence of ancient life on Mars. 

Recently, \citet{Yoshida2023} modeled differences in $^{13}$C between CO and CO$_{2}$ at various heights in the Martian atmosphere. At every measured height, they found that CO$_{2}$ photolysis had caused a depletion in $^{13}$C in CO relative to the degree of fractionation seen in CO$_{2}$. \citet{Yoshida2023} use their carbon isotope ratio measurements to inform on past atmospheric loss. Additionally, changes in the $^{13}$CO$_{2}$ and $^{12}$CO$_{2}$ were measured in the lower atmosphere of Mars over time and altitude with ExoMars Gas Trace Orbiter \citep{Liuzzi2022}. Their results showed that temporal changes in the two most abundant CO$_{2}$ isotopologues largely varied together and likely correspond to changes in the amount of dust rather than seasonal variations. Likewise, the carbon isotope ratio did not show a trend with altitude. With the launch of JWST, $^{13}$CO$_{2}$ will be increasingly commonly measured on other terrestrial solar system bodies such as Pluto and Callisto. Atmospheric measurements of isotope ratios for exoplanets will be limited to bulk atmospheric average measurements initially.

In addition to carbon, sulfur isotopes have also been considered as evidence for life in our solar system \citep{Moreras-Marti2022}. Marine sediments are depleted in heavier sulfur isotopes as the lightest sulfur isotope ($^{32}$S) is preferentially reduced by sulfur-reducing microbes, enriching sulfates (e.g., evaporites) \citep[][and references therein]{Sim2011}. As such, sulfur isotopes may prove to be a valuable biosignature for solar system bodies such as Mars and Europa where \textit{in situ} and sample return measurements will be possible \citep{Moreras-Marti2022}. $\delta\ce{^{34}S}$ has already been measured on Mars as $-0.24\pm0.05$\textperthousand~\citep{Franz2019} and sulfur has been detected on the surface of Europa \citep{Becker2022}. \citet{ChelaFlores2021} posit that sulfur isotope fractionation of $-$70 \textperthousand\ or more is caused by a metabolic process and could be used as evidence for life on solar system bodies. Unfortunately, however, \textit{in situ} geochemical tests on solar system ocean world surfaces in the near future will be challenging \citep[][and references therein]{ChelaFlores2021, Arevalo2019}.

Other gases are also fractionated by life, such as hydrogen and nitrogen. However, there have yet to be studies about their accessibility as biosignatures in our solar system or beyond. Furthermore, when considering isotopes as evidence of metabolic processes outside of the solar system, only metabolically fractionated \textit{gases}---which could potentially alter the spectroscopic atmospheric signature---are relevant. Restricting our analysis to gases severely limits which isotopologues are worthy of assessment as possible bioindicators outside of the solar system. As discussed in Section \ref{intro}, isotopologues of CO$_{2}$ followed by CO are our best potential isotopologue biosignature gas as (1) all known life is carbon-based and carbon remains the best candidate building block of life, no matter its chemical makeup \citep{Petkowski2020}, (2) CO$_{2}$ has a large absorption cross-section, making it readily observable using spectroscopy, and (3) CO$_{2}$ has the largest spectral separation between the first and second most abundant isotopologues among the most common carbon-bearing species for wavelengths relative to JWST. However, while $^{13}$CO$_{2}$ represents our best opportunity of using an isotopologue as a biosignature gas, there are still many challenges which we discuss in Section \ref{challenges}.

\section{Detectability of Isotopes in Exoplanet Atmospheres} \label{exoplanets}
Before we can evaluate the usefulness of isotopologues as biosignature gases in exoplanetary atmospheres, we must understand the current state of the field and what observations are possible with modern instrumentation. At present, we can study exoplanet atmospheres to detect molecules such as water, carbon monoxide, carbon dioxide, hydrogen cyanide, and ammonia. Over the next few years, JWST will provide an unprecedented understanding of the molecular composition of a range of exoplanets. Atmospheric measurements via transmission spectroscopy will not only inform on the composition, but can also be used to detect the presence of clouds and place constraints on the atmospheric temperature-pressure profile. With this knowledge, we can assess the habitability of a planet.

\subsection{Recent Measurements and Simulated Detections}
Today atmospheric signals can be detected remotely through direct imaging, transit spectroscopy, and High-Resolution Cross-Correlation Spectroscopy (HRCCS). In order to detect isotopologues, instruments must be capable of resolving their features spectroscopically. Resolving isotopologues necessitates instruments with a high spectral resolution and large aperture. 

As a prelude to isotopologue detection in the atmospheres of large exoplanets, \citet{Crossfield2019} measured $^{13}$CO spectroscopically in the photospheres of two small, M dwarf stars. \citet{Crossfield2019} found the binary system GJ 745 AB to be considerably depleted in $^{13}$C with a $^{12}$C/$^{13}$C of $296\pm45$ and $224\pm26$ relative to the solar photosphere value of $\sim$80 \citep{Woods2009,Ayres2006}. The authors attribute the enhancement in $^{12}$C to the accretion of supernovae ejecta \citep{Crossfield2019}.

Isotopes have only recently been quantified in an exoplanet atmosphere. Using the ground-based SINFONI instrument on the Very Large Telescope, $^{12}$CO/$^{13}$CO was measured in the atmosphere of super-Jupiter TYC 8998-760-1 b \citep{Zhang2021}. They found that the atmosphere was significantly enriched in the heavier isotope relative to both the local interstellar medium (ISM) ($68\pm15$ \citep{Milam2005}) and solar system ($\sim$ 89 \citep[][and references therein]{Woods2009}) with a ratio of $31^{+17}_{-10}$, which is equivalent to $\delta\ce{^13C}$ of $1900^{+1400}_{-1000}$ \citep[][and references therein]{Zhan2021}. The authors ascribe the enhancement to the accretion of enriched ices during TYC 8998-760-1 b's formation past the CO snowline \citep{Zhang2021}. Ices are enriched in $^{13}$CO due to low temperature isotope exchange reactions in the gas phase, which then freeze the enriched CO into ice past the snowline \citep{Langer1984, Charnley2004, Jorgensen2016}. Similarly, \citet{Line2021} also measured $^{12}$CO/$^{13}$CO in an exoplanet atmosphere, finding a range of 10.2–42.6 at 68\% confidence for hot Jupiter WASP-77 A b. The enrichment in $^{13}$C suggests that WASP-77 A b may have formed past the CO snowline and migrated inward to its current location. 

The detectability of isotopologues in planets smaller than giant exoplanets has also been evaluated. \citet{Glidden2022} modeled observations of $^{13}$CO$_{2}$ in the atmospheres of temperate sub-Neptunes using JWST. Such worlds have recently been proposed as potential hosts to aerial biospheres \citep{Seager2021}. $^{13}$CO$_{2}$ could potentially be used as evidence of an aerial biosphere. As isotopologues of carbon-bearing species have been used as potential evidence for life on Earth, \citet{Glidden2022} sought to evaluate this technique for exoplanet atmospheres. Given their larger size, atmospheric constituents of sub-Neptunes are more readily observable than those of terrestrial planets for two important reasons. First, their size makes the contrast between planet and host star larger during the transit. Second, the larger size leads to more gravitational attraction between the planet and its atmosphere, allowing the planet to retain lighter gases like hydrogen and helium. Lighter gases are important as they lead to larger, puffier atmospheres, which are much easier to detect using transit spectroscopy then atmospheres dominated by heavier gasses like CO$_{2}$ and H$_{2}$O. However, existing theories of abiogenesis predict that life would not emerge in a gaseous exoplanet, as life requires a rocky surface to arise \citep[][and references therein]{Glidden2022}. Thus, the detection of bioindicators on gaseous exoplanets would falsify existing theories of the origin of life, providing a unique opportunity to empirically test whether rocky planetary surfaces are required for abiogenesis. The key parameters that influence detectability of atmospheric spectral features are the magnitude of the host star, the stellar type (size of the star), planet size, scale height (atmospheric composition), and the transit duration. \citet{Glidden2022} found that $^{13}$CO$_{2}$ could only be distinguished from $^{12}$CO$_{2}$ for the most ideal target (cooler, larger sub-Neptune, around a bright, small star with a low mean molecular weight atmosphere), which has yet to be conclusively discovered. Other carbon-bearing isotopologues, such as $^{13}$CH$_{4}$ and $^{13}$CO, will be difficult to distinguish at the resolution of JWST for small planets given the spectral resolution and unknown atmospheric constituents. Current on-going observations with JWST have so far focused on evaluating if terrestrial planets around M dwarfs even have atmospheres using the 4.3$\mu$m feature of CO$_{2}$ and 3.3$\mu$m CH$_{4}$ feature (JWST GO \#1981, PIs: Stevenson \& Lustig-Yaeger). Observations of $^{13}$CO$_{2}$ in potentially habitable planets using ground-based instruments will be challenging for current instruments due to the low contrast between host stars and temperate planets. The future Extremely Large Telescopes will have large apertures, high resolution, and may be able to reach the contrast necessary for temperate worlds \citep{Snellen2013, Rodler2014, Serindag2019}.

\subsection{Why $^{13}$CO$_{2}$ is the Most Detectable Isotopologue from Space} 

With the successful launch of JWST, CO$_{2}$ now presents our best opportunity to distinguish multiple isotopologues of the same gas in an exoplanet atmosphere. CO$_{2}$ is expected to be present in large abundances in exoplanetary atmospheres \citep[][and references therein]{Rustamkulov2023, Molliere2017}, and it has a strong spectral signature, accessible to JWST with several instrument modes. Biological processes also fractionate other elements, such as nitrogen, sulfur, oxygen, hydrogen, and iron \citep[e.g.,][and references therein]{Epstein1988, Canfield2001, Nealson2003, Unkovich2013, Craine2015, Denk2017, Bogard2017, Mooshammer2020, Wieloch2022}. On Earth, biologically mediated isotopologues are often measured in the soil or rock record. Here, we focus on volatiles as they may be detectable in an exoplanet atmosphere.

\begin{figure*}[!ht]
    \centering
    \includegraphics[width=\linewidth]{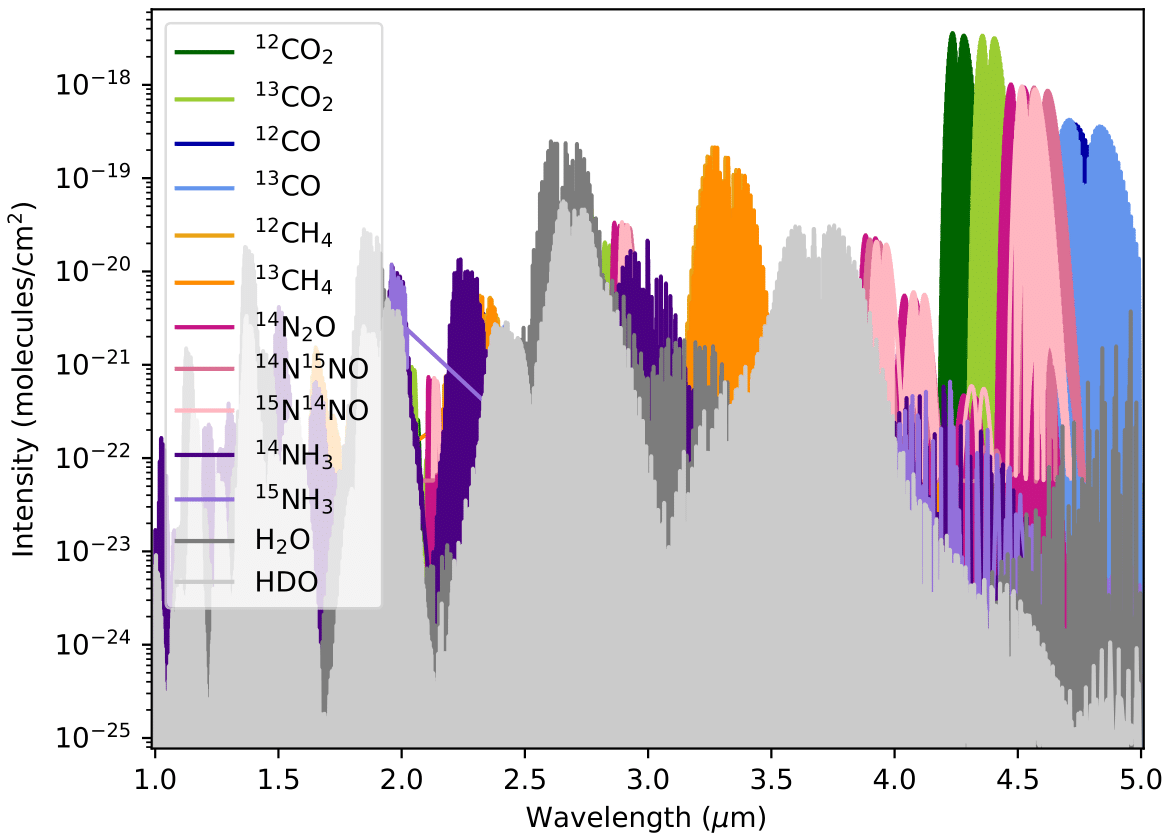}
    \caption{Intensity of key gas isotopologues, which could potentially be metabolically fractionated. The intensity is shown along the y-axis and the wavelength in microns along the x-axis. The wavelength range shown includes the strongest features of each isotopologues relevant for JWST. The $^{13}$CO$_{2}$ and $^{12}$CO$_{2}$ isotopologues (in green) are both the strongest absorbers and the most well spectrally separated, making them the best isotopologue pair for spectroscopic detection with JWST.}
    \label{fig:lines}
\end{figure*}

Figure \ref{fig:lines} shows the intensity of the strongest features of key gas isotopologues of CO$_{2}$, CH$_{4}$, N$_{2}$O, NH$_{3}$, and H$_{2}$O at various wavelengths covered by JWST. The line lists in the figure come from the High-Resolution Transmission Molecular Absorption (HITRAN) database \citep{Gordon2017} and are processed using the HITRAN Application Programming Interface ({\tt HAPI}) \citep{Kochanov2016}. The intensity for each molecule in HITRAN assumes a temperature of 296 K \citep{Rothman1996}. For each isotopologue, we normalize the intensity by its terrestrial abundance to account for the weighting used by HITRAN \citep{Gordon2022, DeBievre1984}. CO$_{2}$ is clearly the strongest absorber and has the largest spectral separation between its first and second most abundant isotopologue. Given its shape and bond strength, the CO$_{2}$ molecule is more altered by the additional neutron than CO and CH$_{4}$. The relatively large shift between $^{13}$CO$_{2}$ and $^{12}$CO$_{2}$ makes the isotopologues most likely to be accessible to instruments such as JWST. Within the JWST wavelength coverage, the two most abundant isotopologues of CO$_{2}$ are maximally separated by 0.121$\mu$m compared with 0.105$\mu$m for CO and 0.010$\mu$m for CH$_{4}$.

The bond strength of the molecule influences how much the additional neutron affects the isotopologues' spectral signature. Differences between $^{13}$CO$_{2}$ and $^{12}$CO$_{2}$ can be attributed to the addition of a single neutron to the nucleus of the carbon atom. CO$_{2}$ is a small, linear molecule. Adding an additional neutron to the carbon atom does not impact the symmetry of the molecule. Imagining the molecule as masses connected by springs, we can understand that increasing the mass of the central molecule will increase the reduced mass of the molecule, thereby decreasing the vibrational energy levels. This is well illustrated by the asymmetric stretch band in CO$_{2}$ around 4.3 $\mu$m, as shown in Figure \ref{fig:lines}.

\begin{figure*}[!ht]
    \centering
    \includegraphics[width=\linewidth]{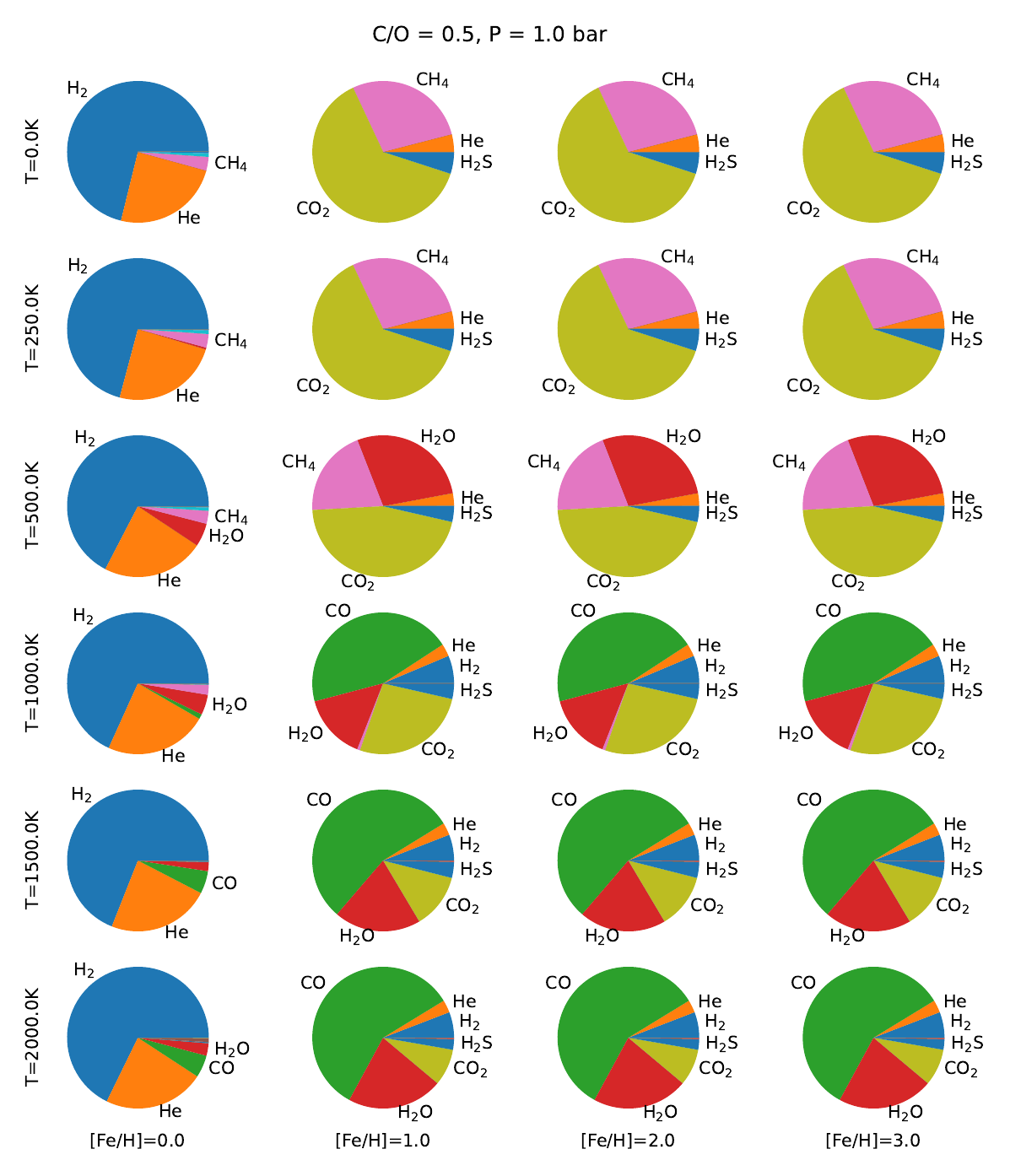}
    \caption{Relative mass fraction composition of planetary atmospheres assuming chemical equilibrium. Each pie plot is colored according to molecular composition for C/O = 0.5 (solar value) and pressure of 1 bar (Earth's surface value) \citep{Madhusudhan2012}. Each column corresponds to a metallicity ([Fe/H]), as given along the bottom of the figure. Each row corresponds to a temperature, labeled along the left side. For temperate values, CO$_{2}$ and CH$_{4}$ dominate over CO in all cases. For carbon-bearing species, CO$_{2}$ dominates for low C/O and CH$_{4}$ dominates for higher C/O.}
    \label{fig:pie}
\end{figure*}

The detectability of a given isotopologue is also dependent on its abundance within the exoplanet's atmosphere. Figure \ref{fig:pie} shows the relative mass fractions of gases in planetary atmospheres, assuming chemical equilibrium with C/O of 0.5 and a pressure of 1 bar for a range or temperatures and metallicities, as calculated using {\tt petitRADTRANS} \citep{Molliere2019prt}. Temperature increases down the rows, while metallicity increases across the columns. For temperate atmospheres, CO$_{2}$ dominates for low C/O ratios, while CH$_{4}$ dominates for higher C/O ratios.

\subsection{Formation of Carbon Isotopes}
For completeness, we discuss where carbon isotopes in planet-forming materials originate. Nucleosynthesis in stars is responsible for the formation of carbon isotopes. In particular, $^{12}$C is produced through helium burning using the triple-alpha process while the carbon--nitrogen--oxygen cycle of hydrogen burning also synthesizes both $^{12}$C and $^{13}$C. Asymptotic giant branch (AGB) stars are largely responsible for the production of carbon \citep{Kobayashi2011}. During the third dredge-up, deeper products of nuclear fusion are brought to the surface through convection \citep{Kobayashi2011}. The outer envelope of the star becomes enriched in carbon, which can be jettisoned into the surrounding ISM and eventually used to form new stars and planets. The mass of the AGB star affects if $^{12}$C or $^{13}$C is produced. $^{12}$C is generally fused in stars of $1-4$ M$_{\odot}$, while $^{13}$C is largely created in intermediate mass stars of $4-7$ M$_{\odot}$ \citep{Kobayashi2011}. Carbon is also synthesized though core-collapse supernovae at a similar yield to AGB stars.

When a newborn star is formed from a collapsing molecular cloud, the carbon isotope composition of the new solar system is tied to the ratio of the star-forming region. Areas with more low-mass AGB stars have higher $^{12}$C/ $^{13}$C than areas with intermediate-mass AGB stars \citep{Kobayashi2011}. However, $^{12}$C/ $^{13}$C is not homogenized throughout a protoplanetary disk \citep{Woods2009}. After formation, carbon can further fractionate through two abiotic processes: chemical exchange reactions and photodissociation \citep{Woods2009}. The dominant isotope exchange reaction is $^{13}$C$^{+} + ^{12}$CO $\rightleftharpoons ^{13}$CO$+ ^{12}$C$^{+} + \Delta E$ \citep{Watson1976, SmithAdams1980, Woods2009}. Photodissociation more strongly impacts $^{13}$CO as it is relatively less shielded than $^{12}$CO \citep{Sheffer1992, Woods2009}. Thus, the $^{12}$C/ $^{13}$C ratio differs across the disk, largely dependent on temperature differences within the disk. Modeling suggests that carbon isotope ratios should vary radially and axially within a protoplanetary disk \citep{WoodsWillacy2009}. Recently, \citet{Yoshida2022} were able to measure radial variations in $^{12}$C/$^{13}$C in a protoplanetary disk for the first time. They found differences based on both molecular species and radial location within the disk.

\section{Major Challenges with Using Isotopologues as Biosignature Gases} \label{challenges}

Despite the increasing ease of detecting isotopologues with new and near-future instruments, there are several major challenges with their use as biosignature gases. Here, we outline the three most significant challenges.

\subsection{Establishing an Abiotic Baseline for $^{12}$C/$^{13}$C in Exoplanet Atmospheres}\label{baseline}

A baseline value for $^{12}$C/$^{13}$C is essential to be able to compare against any measured values of $^{13}$C or $^{12}$C to determine if an enhancement in atmospheric $^{13}$C can be attributed to a metabolic process, such as photosynthesis. Planetary isotope ratios should be similar to those of their host star and local ISM. Thus, the host star or the local ISM could be used to roughly benchmark the abiotic isotope ratio of the exoplanet's atmosphere \citep{MolliereSnellen2019}. Carbon isotope ratios from exocomets could also be used, but such measurements are far beyond our detection capabilities. 

If we are to make inroads measuring deviations in $^{12}$C/$^{13}$C in planetary atmospheres, we must first make isotopologue measurements (using $^{12}$CO/$^{13}$CO) in host stars. The striking difference between $^{12}$C/$^{13}$C in M dwarfs GJ 745 AB ($296\pm45$ and $224\pm26$ \citep{Crossfield2019})  and our Sun ($93.5\pm3$ \citep{Lyons2018}) underscores the importance of spectroscopically measuring host stars to determine if the enhancement seen in GJ 745 AB is specific to that particular region of the Galaxy (due to enrichment from neighboring supernovae) or a more generalized characteristic of M dwarf stars. Furthermore,  $^{12}$C/$^{13}$C was measured in seven young stellar objects (YSOs) to be between $\sim$85 and 165 \citep{Smith2015}. The YSOs were more depleted in $^{13}$C compared with our solar system ($\sim$ 89 \citep[][and references therein]{Woods2009}) and local ISM ($68\pm15$, \citet{Milam2005}). Current models of galactic chemical evolution do not explain this discrepancy \citep{Smith2015}. Stellar measurements of carbon isotope ratios are especially important for better refining our understanding of galactic evolution and isotope enrichment as well as setting a benchmark for isotopologue measurements in exoplanet atmospheres. However, as carbon isotopes have only been measured in two M dwarfs, the local ISM could also be used as a proxy until better stellar spectral measurements are obtained.

\begin{figure*}[!ht]
    \centering
    \includegraphics[width=\textwidth]{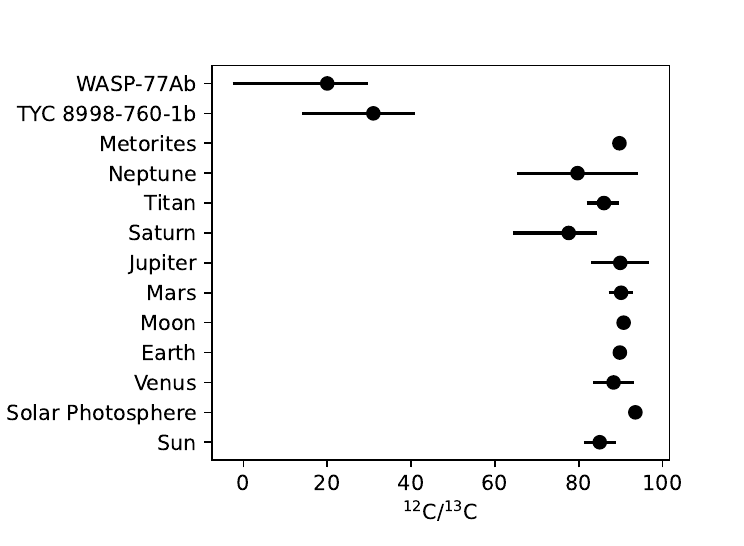} 
    \caption{Carbon isotope ratios for the solar system (average values from \citet[][and references therein]{Woods2009} and solar photosphere value from \citet{Lyons2018}) and the two measured exoplanet atmospheres \citep{Zhang2021, Line2021}. The object name is shown on the y-axis and the carbon isotope ratio is shown on the x-axis. There is a clear clustering of solar system carbon isotope ratios around $\sim$ 89 \citep{Woods2009}. The two measured carbon isotope ratios for exoplanets both show an enrichment in $^{13}$C, likely attributed by ice accretion. While TYC 8998-760-1b is located far ($<$ 160 AU) from its host star, WASP-77 A b is a close-in, hot Jupiter. The enrichment in $^{13}$CO suggests that WASP-77 A b may have formed past the CO snowline and migrated to its current location.}
    \label{fig:ratios}
\end{figure*}

However, even if we could measure carbon isotope ratios in host stars, we would still be unable to use remote atmospheric measurements of carbon isotope ratios as strong evidence of life. Consider our own solar system where $^{12}$C/$^{13}$C has been measured extensively for the Sun, Venus, Earth, the Moon, Mars, Jupiter, Saturn, Titan, Neptune, and many meteorites and comets \citep[e.g.,][and references therein]{Lyons2018, Woods2009}. Since the beginning of remote observations, significant progress has been made in the error estimation of carbon isotope ratios, in particular those of the Sun. While subtle differences between solar system bodies have been used to suss out their formation histories, the bulk carbon isotope ratios measured across the various solar system bodies are largely homogeneous. As shown in Table \ref{tab:ssratios} and Figure \ref{fig:ratios}, the $^{12}$C/$^{13}$C ratios are generally consistent within the solar system. For example, the mean value of Venus is $88.3\pm4.9$; the mean value for Earth is $89.8\pm1.4$; the mean value for Mars is $90.1^{+3.0}_{-2.9}$, and the mean value of Jupiter is $89.9^{+6.9}_{-6.8}$ \citep{Woods2009}. 

Given these values, it is clear that (1) as measured in the bulk, carbon isotope ratios are largely similar across different solar system bodies, and (2) Earth does not stand out as a living world---within the solar system---despite billions of years of metabolic activity reprocessing carbon. Life only drives small ($\sim$ 0.5 to 7\%) changes in carbon isotope composition, which will be next to impossible to detect remotely with enough precision in a terrestrial exoplanet atmosphere. 

\begin{deluxetable}{cc}
\tabletypesize{\small}
\tablecolumns{2}
\tablewidth{0pt}
\tablecaption{Bulk Carbon Isotope Compositions for Solar System Bodies} \label{table:Table1}
\tablehead{\colhead{Object}	& \colhead{$^{12}$C/$^{13}$C}}
\startdata
Sun\textsuperscript{1}		& $93.5\pm3.0$\\	
Venus\textsuperscript{2}		& $88.3\pm4.9$\\
Earth\textsuperscript{2}		& $89.8\pm1.4$\\
Moon\textsuperscript{2}		& $90.7\pm0.6$\\
Mars\textsuperscript{2}		& $90.1^{+3.0}_{-2.9}$\\
Terrestrial\textsuperscript{2}		& $89.7\pm1.5$\\
Jupiter\textsuperscript{2}		& $89.9^{+6.9}_{-6.8}$\\
Saturn\textsuperscript{2}		& $77.6^{+13.3}_{-6.8}$\\
Neptune\textsuperscript{2}		& $79.7\pm14.5$\\
TYC 8998-760-1 b \textsuperscript{3} & $31^{+17}_{-10}$\\
WASP-77 A b \textsuperscript{4} & 10.2–42.6 at 68\% confidence\\
GJ 745 A \textsuperscript{5} & $296\pm45$ \\ 
GJ 745 B \textsuperscript{5} & $224\pm26$\\ 
\enddata
\label{tab:ssratios}
\tablecomments{\footnotesize\textsuperscript{1} \citet{Lyons2018}. {\textsuperscript{2} Mean values from compiled list of sources within  \citet{Woods2009}}. For context, we also include carbon isotope ratios for exoplanets \footnotesize\textsuperscript{3} \citet{Zhang2021} and \footnotesize\textsuperscript{4} \citet{Line2021} and for two M dwarf stars \textsuperscript{5}\citet{Crossfield2019}.}
\end{deluxetable}

\subsection{Disentangling Abiotic and Biotic Isotope Fractionation}\label{abiotic}

Another difficulty with using isotopologues as biosignature gases is disentangling abiotic and biotic isotope fractionation. Carbon isotope fractionation can occur through several metabolic processes. Photosynthesis is the most well known, but other metabolisms such as chemosynthesis also cause fractionation \citep{Hayes2001, Zyakun2009, Havig2011}. Photosynthesis is broadly accepted as likely to occur on other planets with life \citep[e.g.,][]{Seager2005,Kiang2007}. Biotic carbon isotope fractionation can occur on Earth when plants, algae, and cyanobacteria photosynthesize through two main processes \citep{Still2017}. First, the lighter $^{12}$CO$_{2}$ more readily diffuses through the stomata on plant leaves \citep{Still2017}. Secondly, the Rubisco enzyme used in photosynthesis preferentially removes $^{12}$C as it binds more efficiently than $^{13}$C to one of the enzyme's active sites due to the kinetic isotope effect \citep{McNevin2007}. The magnitude of carbon isotope fractionation is quite small ($\sim1-4\%$) and differs between different photosynthetic pathways \citep[e.g.,][and references therein]{OLeary1988, Kirkels2022}. 

There are many abiotic processes that fractionate carbon. Abiotic processes can lead to false positive detections or potentially erase biological fractionation. For example, carbon isotope fractionation can be caused through volcanism \citep{Mattey1991,Bada2023VolcanicEon}. The degassing of basaltic magma enriches the melt in $^{12}$CO$_{2}$, while enhancing the gas in $^{13}$CO$_{2}$ \citep{Mattey1991}. The magnitude of this effect is only $\sim2$\textperthousand\ \citep{Mattey1991}. In addition, \citet{Ricci2023} reported $^{13}$C and D-depleted abotic methane from high temperate volcanic gas. Volcanism is of particular interest as it may be necessary for the origin of life. Our only example of a habitable planet (Earth) is subject to volcanic activity. Thus, carbon isotope fractionation through magma outgassing will likely prove an important false positive that will need to be disentangled from biotic fractionation in exoplanet atmospheres. As volcanism may be a necessary component of the ``recipe for life" \citep[e.g.,][]{Leman2004}, this may confound the use of carbon isotope composition as a bioindicator. 

\subsection{Limited Number of Known and Anticipated Targets}
\begin{figure*}
    \centering
    \includegraphics[width=\textwidth]{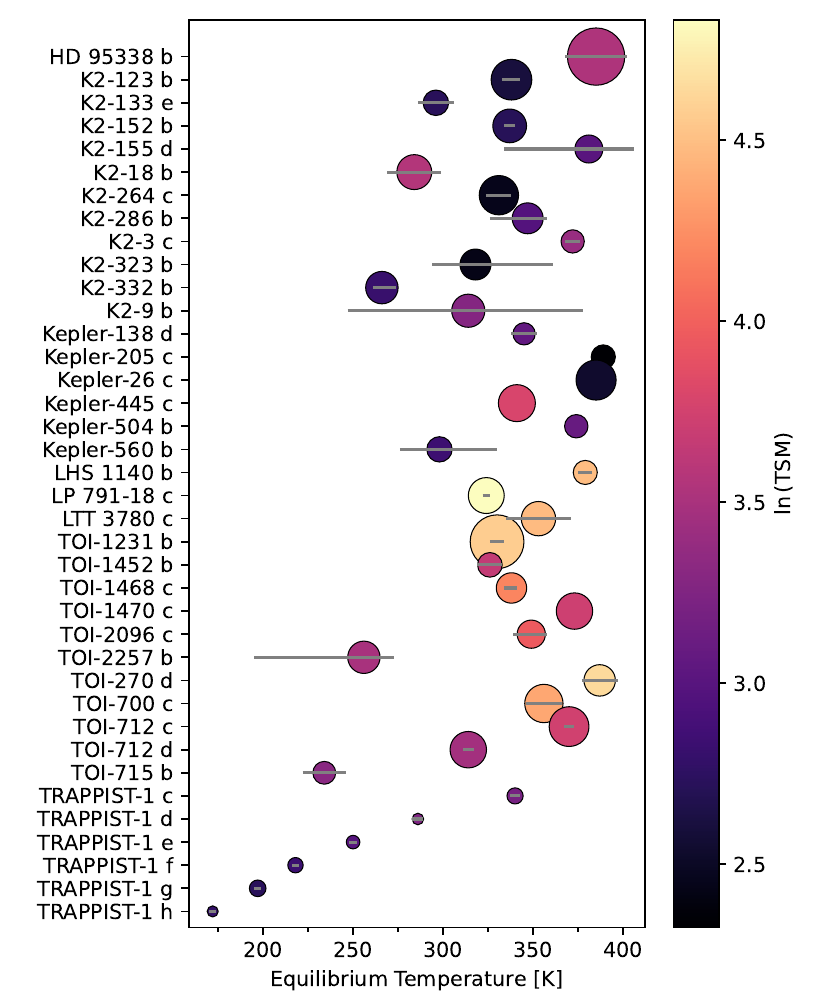} 
    \caption{Cool (T $<$ 389 K) terrestrial and sub-Neptune planets (R$_{p} < $ 4 R$_{\oplus}$) shown along the x-axis in temperature and the y-axis by alphabetical order. Marker size is relative to planet size. Marker color is coded to the ln(TSM). There are very few small, cool planets with a high TSM. The best temperate terrestrial candidates by TSM are TRAPPIST-1~c, d, e, f, and g. The highest TSM small sub-Neptunes (1.5 R$_{\oplus}$ $<$ R$_{p}$ $<$ 2.75 R$_{\oplus}$) candidates are LP 791-18 c and TOI-270 d. The highest TSM of cool planets between from 2.75 R$_{\oplus} <$ R$_{p}$ $<$ 4 R$_{\oplus}$ are Kepler-51~d and TOI-1231~b.}
    \label{fig:targets}
\end{figure*}

To detect isotopologues, we need not only the right instrument, but also the right targets. There are few suitable temperate (habitable zone) targets around small, bright stars. There are only approximately 20 known temperate terrestrial planets and 49 temperate sub-Neptunes. Of these potential candidates, most have a low Transit Spectroscopy Metric (TSM) \citep{Kempton2018}. The TSM is calculated using the radius of the planet and host star, the mass of the planet, the equilibrium temperature of the planet, and the magnitude of the target. The TSM approximates the expected signal of the target's spectral features, assuming the atmosphere is clear (a large caveat). To be a good candidate for atmospheric transmission spectroscopy, terrestrial planets (R$_{p} < 1.5$ R$_{\oplus}$) should have a TSM $>$ $\sim 10$ (ln(TSM) $>$ $\sim 2.3$), while sub-Neptune planets (4 R$_{\oplus}$ $<$ R$_{p}$ $<$ 10 R$_{\oplus}$) should have a TSM of at least $>$ $\sim 90$ (ln(TSM) $>$ $\sim 4.5$) to be a good candidate \citep{Kempton2018}. Of the temperate small planets known to date, 10 have a ln(TSM) $>$2 and only 3 (K2-18 b, TOI-2257 b, and TRAPPIST-1 d) have a ln(TSM) $>$ 3. Figure \ref{fig:targets} shows known temperate small planets. Temperature is on the x-axis, planet name is on the y-axis, and ln(TSM) dictates the color. 

The Transiting Exoplanet Survey Satellite (TESS) mission has finished its nearly all sky survey to look for transiting planets orbiting our nearest and brightest stellar neighbors \citep{Ricker2014}. Considered the ``finder scope" for JWST follow-up observations, TESS continues to look for more promising candidates during its ongoing extended mission (currently in mission year 5). So far, there are 6,213 TESS Targets of Interest, 282 confirmed planets, 151 of which with a radius, R $<$ 4 R${\oplus}$, where R${\oplus}$ is the radius of the Earth (NASA). It is likely that some TESS planet candidates larger than a super-Earth will be favorable for future detections of CO$_{2}$ isotopologues. The longer TESS observes the sky during its extended mission, the higher chance we have of detecting longer period, and thus cooler, planets. Additionally, the longer integration time can also boost the signal-to-noise of a planet's transit depth, potentially revealing more small planet candidates. Recent simulations predict that TESS will find $601\pm44$ planets with R $<$ 2 R$_{\oplus}$ and $3027\pm202$ planets between 2 R$_{\oplus} <$ R $<$ 4 R$_{\oplus}$ during its primary and first two extended missions ($\sim7$ years of observations) \citep{Kunimoto2022Predicting17}.

\section{Carbon Isotope Ratios as Supportive Evidence for Life} \label{solutions}

Despite the many difficulties associated with using carbon isotope ratios as unambiguous evidence for life, there are still strategies for using measurements of carbon isotopes as supportive evidence for the presence of life. Indeed, \citet{Rothschild1989} concluded that carbon isotope ratios could only be used as supportive evidence for past metabolic activities even on Mars due to (1) the number of false positive scenarios and (2) the relatively large range of isotope fractionation possible due to different types of metabolic activity. Here, we outline possible solutions for exoplanets to mitigate our uncertainty that any detected fractionation could be abiotic in origin. 

\subsection{Seasonality of Carbon Isotope Fractionation: A Possible Solution}

The Earth's atmospheric carbon isotopic composition varies seasonally \citep{Keeling2000, Still2017}. During the warmer growing season, $^{13}$CO$_{2}$ increases in Earth's atmosphere relative to $^{12}$CO$_{2}$ as plants grow and photosynthesize more, preferentially using $^{12}$C to build more plant material. The reverse happens during the winter months when photosynthesis decreases. Periodic changes in the carbon isotope ratio can also be seen in the photosynthetic day-night cycle on Earth \citep{Keeling2000, Still2017}.

Thus, a possible way to rule out the abiotic fractionation of carbon on remote planets is to look for oscillations in the $^{13}$C/$^{12}$C ratio that corresponds to the planet's seasonal cycle. If seasonal variations can be seen, it would strongly increase our confidence that any observed carbon isotope fractionation was driven by a metabolic rather than a planetary process. However, as we are unable to even distinguish carbon dioxide isotopologues in the atmospheres of terrestrial planets with upcoming telescopes, detecting seasonal changes will remain well out of reach for decades to come. While seasonal variations in atmospheric carbon isotope ratios on Earth are just $\sim$1\textperthousand~(according to data from NOAA's Global Monitoring Lab \citep{NOAAGlobalMonitoringLab}), the magnitude of seasonality could be larger for a planet with less CO$_{2}$ and higher productivity. While we will not be able to measure seasonality for carbon isotope ratios, observing the seasonality of gases produced through photosynthesis would increase our confidence in the likelihood that a planet hosts life. Freeze--thaw cycles can also lead to seasonal changes in atmospheric gases through both abiotic and biotic processes \citep[i.e.,][]{Kurganova2015}.

\citet{Olson2018} evaluated how seasonality in O$_{2}$, CO$_{2}$, and other gases could be used to rule out false positive detections. As the transit method intrinsically selects for planets at shorter periods, many known transiting planets are tidally locked with their host star. Tidally locked planets lack a day--night cycle and also lack the seasonal cycle caused by a tilt in the spin axis relative to the orbital plane. Only the eccentricity (if any) of their orbits would cause seasonality. In addition, the viewing angle of a transiting planet will limit our ability to view different seasons \citep{Meadows2018, Olson2018}. Thus, seasonality will not be expected in most transiting planets. Additionally, seasonality will be best observed through direct imaging as the system can be viewed face-on. It will be difficult, if not impossible, to perform follow-up observations of transiting planets with direct imaging to look for seasonality given the edge-on transit geometry. 

\subsection{The Case for Observing Carbon Isotope Ratios in Non-Terrestrial Planet Atmospheres: Isotopes as a Tool to Discern Formation Mechanisms}\label{planetarysystems}

Giant planets offer a proving ground for adding isotopologue detections to our spectroscopic toolkit. We can use the observations of giant planets to test our theories and refine our modeling capabilities. In addition, as the detection of carbon isotope ratios in non-terrestrial planets would not be due to life, any detection would help us to establish a baseline for false positive detection of carbon isotope ratios as a biosignature. Furthermore, isotopes can be used to unravel exoplanet mysteries unrelated to the search for life. The isotopic composition of a planetary system is originally inherited from the molecular cloud out of which it formed \citep[][and references therein]{Nomura2022TheSystem}. As such, understanding what process produces different isotope fractionation and how it varies on a galactic scale is important for providing context for future measurements of isotopologues as biosignature gases. 

Temperature often dictates which isotope fractionation reaction dominates \citep{Nomura2022TheSystem}. Different reactions are possible in the gas and solid phase and within different temperature regimes within these phases. In the gas phase, isotope fractionation is caused by isotope exchange reactions and isotope selective photodissociation \citep[e.g.,][and references therein]{Nomura2022TheSystem}. At their freezing point, enriched gases can freeze onto the surface of dust grains as ice \citep[e.g.,][]{BrownMillar1989, YurimotoKuramoto2004, Nomura2022TheSystem}. Within the disk, gas pressure gradients are responsible for the redistribution and migration of enriched species \citep{Nomura2022TheSystem}. Therefore, it may be possible to use the isotopic composition of a planetary body to work out where it formed in the protoplanetary disk as different regions are subject to different temperature regimes. For example, it has been hypothesized that $^{12}$CO/$^{13}$CO can be used to assess if a planet formed via ice-accretion past the CO snowline or via gas-accretion interior to the CO snowline \citep{Zhang2021}. Atmospheres enriched in $^{13}$CO are likely formed through ice-accretion past the snowline. If planets with high $^{13}$CO are found close to their host star, it may be evidence of migration.

Recently, JWST has been used to measure $^{13}$CO$_{2}$ for both a protoplanetary disk \citep{Grant2023} and molecular cloud ices \citep{McClure2023}. The observation of $^{13}$CO$_{2}$ in the GW Lup disk marks the first time this isotopologue has been measured in a protoplanetary disk, the birthplace of future exoplanets. Expanding on these observations is key for using $^{13}$CO$_{2}$ as a tracer of planet formation, from the origin of planetary materials in a molecular cloud through the formation of a planetesimal up to the final stage of planet formation. The pioneering work of \citet{Grant2023} and \citet{McClure2023} lays the foundation for what we will be able to learn with JWST about exoplanet formation through the lens of isotopologue measurements.

\section{Conclusions} \label{conclusion}

Isotopologues are unlikely to prove fruitful biosignature gases in exoplanetary atmospheres in the foreseeable future. While metabolically mediated carbon isotope fractionation is the most readily detectable using $^{13}$CO$_{2}$ due to CO$_{2}$'s significant spectral separation and large absorption cross-section, followed closely by CO, there are still significant challenges associated with using carbon isotope ratios as a bioindicator outside of the solar system. CO$_{2}$ isotopologues are most accessible to JWST for cool, large planets around small, bright stars. However, the number of such targets is extremely limited. For smaller, potentially habitable sub-Neptunes and terrestrial planets, isotopologues are only accessible for the most idealized targets with large scale heights. While we have focused largely on CO$_{2}$, the key challenges associated with using carbon isotope ratios as biosignature gases in exoplanet atmospheres are universal to other species. Carbon isotope ratios in exoplanet atmospheres will need to be compared with a known abiotic reservoir to look for changes that may be caused by life. Baseline carbon isotope ratio values in other exoplanetary systems are yet to be established. While the ISM and host star could serve as a proxy, life only causes small changes in isotope ratio values and different metabolic processes cause varied amounts of fractionation. As \citet{Rothschild1989} found true for Mars in 1989, carbon isotope data will only ever prove itself to be supportive evidence for life on exoplanets. Thus, isotopologues as biosignature gases are best left as a solar system endeavor where values can be precisely measured \emph{in situ} over a range of surface locations. 

However, isotopologues are still worth detecting in exoplanet atmospheres as they inform on planetary formation history and evolution. Within a protoplanetary disk, isotope composition varies both radially and axially \citep{WoodsWillacy2009}. Carbon isotope ratios are believed to change corresponding to snowlines \citep{Zhan2021}. Thus, measuring carbon isotope ratios in an exoplanet atmosphere may be able to provide clues about the chemical environment in which the atmosphere formed and evolved. In particular, ice accretion may drive $^{13}$CO enrichment and thus, if enriched planets are found close to their host star, they likely migrated to that location rather than forming \textit{in situ}. Thus, isotopologue measurements outside of the solar system should focus on sussing out formation location and evolutionary processes rather than being used as evidence of microbial activity. 

\vspace{6pt} 

This work was partially funded by NASA grant 80NSSC20K0586. A.G. acknowledges support from the Robert R. Shrock Graduate Fellowship. We would like to thank Kaitlin Rasmussen for a productive conversation about ground-based observations and Robert Hargreaves for sharing his expertise on spectral line lists. Furthermore, we would like to thank the anonymous reviewers for their comments, which greatly enhanced the completeness of this work. We also thank the special issue editors, Dr. Amy E. Hofmann and Prof. John M. Eiler, for their insightful and valuable comments on isotope biogeochemistry, which significantly improved the manuscript.

\bibliography{references.bib}{}

\begin{thebibliography}{}
\expandafter\ifx\csname natexlab\endcsname\relax\def\natexlab#1{#1}\fi
\providecommand{\url}[1]{\href{#1}{#1}}
\providecommand{\dodoi}[1]{doi:~\href{http://doi.org/#1}{\nolinkurl{#1}}}
\providecommand{\doeprint}[1]{\href{http://ascl.net/#1}{\nolinkurl{http://ascl.net/#1}}}
\providecommand{\doarXiv}[1]{\href{https://arxiv.org/abs/#1}{\nolinkurl{https://arxiv.org/abs/#1}}}

\bibitem[{Ahrer {et~al.}(2023)Ahrer, Alderson, Batalha, Batalha, Bean, Beatty, Bell, Benneke, Berta-Thompson, Carter, Crossfield, Espinoza, Feinstein, Fortney, Gibson, Goyal, Kempton, Kirk, Kreidberg, L{\'{o}}pez-Morales, Line, Lothringer, Moran, Mukherjee, Ohno, Parmentier, Piaulet, Rustamkulov, Schlawin, Sing, Stevenson, Wakeford, Allen, Birkmann, Brande, Crouzet, Cubillos, Damiano, D{\'{e}}sert, Gao, Harrington, Hu, Kendrew, Knutson, Lagage, Leconte, Lendl, MacDonald, May, Miguel, Molaverdikhani, Moses, Murray, Nehring, Nikolov, Petit dit de~la Roche, Radica, Roy, Stassun, Taylor, Waalkes, Wachiraphan, Welbanks, Wheatley, Aggarwal, Alam, Banerjee, Barstow, Blecic, Casewell, Changeat, Chubb, Col{\'{o}}n, Coulombe, Daylan, de~Val-Borro, Decin, Dos~Santos, Flagg, France, Fu, Garc{\'{i}}a~Mu{\~{n}}oz, Gizis, Glidden, Grant, Heng, Henning, Hong, Inglis, Iro, Kataria, Komacek, Krick, Lee, Lewis, Lillo-Box, Lustig-Yaeger, Mancini, Mandell, Mansfield, Marley, Mikal-Evans, Morello, Nixon, Ortiz~Ceballos, Piette,
  Powell, Rackham, Ramos-Rosado, Rauscher, Redfield, Rogers, Roman, Roudier, Scarsdale, Shkolnik, Southworth, Spake, Steinrueck, Tan, Teske, Tremblin, Tsai, Tucker, Turner, Valenti, Venot, Waldmann, Wallack, Zhang, \& Zieba}]{Ahrer2023}
Ahrer, E.~M., Alderson, L., Batalha, N.~M., {et~al.} 2023, Nature, 614, 649, \dodoi{10.1038/s41586-022-05269-w}

\bibitem[{Alday {et~al.}(2021)Alday, Wilson, Irwin, Trokhimovskiy, Montmessin, Fedorova, Belyaev, Olsen, Korablev, Lef{\`{e}}vre, Braude, Baggio, Patrakeev, \& Shakun}]{Alday2021}
Alday, J., Wilson, C.~F., Irwin, P.~G., {et~al.} 2021, Journal of Geophysical Research: Planets, 126, \dodoi{10.1029/2021JE006992}

\bibitem[{Arevalo {et~al.}(2020)Arevalo, Ni, \& Danell}]{Arevalo2019}
Arevalo, R., Ni, Z., \& Danell, R.~M. 2020, Journal of Mass Spectrometry, 55, \dodoi{10.1002/jms.4454}

\bibitem[{Ayres {et~al.}(2006)Ayres, Plymate, \& Keller}]{Ayres2006}
Ayres, T.~R., Plymate, C., \& Keller, C.~U. 2006, The Astrophysical Journal Supplement Series, 165, 618, \dodoi{10.1086/504847}

\bibitem[{Bada(2023)}]{Bada2023VolcanicEon}
Bada, J.~L. 2023, Nature Communications, 14, 2011, \dodoi{10.1038/s41467-023-37894-y}

\bibitem[{Becker {et~al.}(2022)Becker, Trumbo, Molyneux, Retherford, Hendrix, Roth, Raut, Alday, \& McGrath}]{Becker2022}
Becker, T.~M., Trumbo, S.~K., Molyneux, P.~M., {et~al.} 2022, Planetary Science Journal, 3, \dodoi{10.3847/PSJ/ac69eb}

\bibitem[{Bell {et~al.}(2015)Bell, Boehnke, Harrison, \& Mao}]{Bell2015}
Bell, E.~A., Boehnke, P., Harrison, T.~M., \& Mao, W.~L. 2015, Proceedings of the National Academy of Sciences of the United States of America, 112, 14518, \dodoi{10.1073/pnas.1517557112}

\bibitem[{Bogard {et~al.}(2017)Bogard, Vachon, St.-Gelais, \& del Giorgio}]{Bogard2017}
Bogard, M.~J., Vachon, D., St.-Gelais, N.~F., \& del Giorgio, P.~A. 2017, Biogeochemistry, 133, 347, \dodoi{10.1007/s10533-017-0338-5}

\bibitem[{Brown \& Millar(1989)}]{BrownMillar1989}
Brown, P.~D., \& Millar, T.~J. 1989, MNRAS, 237, 661, \dodoi{10.1093/mnras/237.3.661}

\bibitem[{Canfield(2001)}]{Canfield2001}
Canfield, D.~E. 2001, Reviews in Mineralogy and Geochemistry, 43, 607, \dodoi{10.2138/gsrmg.43.1.607}

\bibitem[{Charnley {et~al.}(2004)Charnley, Ehrenfreund, Millar, Boogert, Markwick, Butner, Ruiterkamp, \& Rodgers}]{Charnley2004}
Charnley, S.~B., Ehrenfreund, P., Millar, T.~J., {et~al.} 2004, Mon. Not. R. Astron. Soc, 347, 157, \dodoi{10.1111/j.1365-2966.2004.07188.x}

\bibitem[{Chela-Flores(2021)}]{ChelaFlores2021}
Chela-Flores, J. 2021, Frontiers in Space Technologies, 2, \dodoi{10.3389/frspt.2021.703809}

\bibitem[{Craine {et~al.}(2015)Craine, Brookshire, Cramer, Hasselquist, Koba, Marin-Spiotta, \& Wang}]{Craine2015}
Craine, J.~M., Brookshire, E.~N., Cramer, M.~D., {et~al.} 2015, Plant and Soil, 396, 1, \dodoi{10.1007/s11104-015-2542-1}

\bibitem[{Crossfield {et~al.}(2019)Crossfield, Waalkes, Newton, Narita, Muirhead, Ment, Matthews, Kraus, Kostov, Kosiarek, Kane, Isaacson, Halverson, Gonzales, Everett, Dragomir, Collins, Chontos, Berardo, Winters, Winn, Scott, Rojas-Ayala, Rizzuto, Petigura, Peterson, Mocnik, Mikal-Evans, Mehrle, Matson, Kuzuhara, Irwin, Huber, Huang, Howell, Howard, Hirano, Fulton, Dupuy, Dressing, Dalba, Charbonneau, Burt, Berta-Thompson, Benneke, Watanabe, Twicken, Tamura, Schlieder, Seager, Rose, Ricker, Quintana, L{\'{e}}pine, Latham, Kotani, Jenkins, Hori, Colon, \& Caldwell}]{Crossfield2019}
Crossfield, I. J.~M., Waalkes, W., Newton, E.~R., {et~al.} 2019, The Astrophysical Journal Letters, 883, L16, \dodoi{10.3847/2041-8213/ab3d30}

\bibitem[{De~Bi{\'{e}}vre {et~al.}(1984)De~Bi{\'{e}}vre, Gallet, Holden, \& Barnes}]{DeBievre1984}
De~Bi{\'{e}}vre, P., Gallet, M., Holden, N.~E., \& Barnes, I.~L. 1984, Journal of Physical and Chemical Reference Data, 13, 809, \dodoi{10.1063/1.555720}

\bibitem[{Denk {et~al.}(2017)Denk, Mohn, Decock, Lewicka-Szczebak, Harris, Butterbach-Bahl, Kiese, \& Wolf}]{Denk2017}
Denk, T. R.~A., Mohn, J., Decock, C., {et~al.} 2017, Soil Biology and Biochemistry, 105, 121, \dodoi{10.1016/j.soilbio.2016.11.015}

\bibitem[{Eiler {et~al.}(1997)Eiler, Mojzsis, \& Arrhenius}]{Eiler1997}
Eiler, J.~M., Mojzsis, S.~J., \& Arrhenius, G. 1997, Nature, 386, 665, \dodoi{10.1038/386665a0}

\bibitem[{Epstein \& Zeiri(1988)}]{Epstein1988}
Epstein, S., \& Zeiri, L. 1988, Proc. Natl. Acad. Sci. USA, 85, 1727, \dodoi{10.1073/pnas.85.6.1727}

\bibitem[{Franz {et~al.}(2019)Franz, Wu, Farquhar, \& Irving}]{Franz2019}
Franz, H.~B., Wu, N., Farquhar, J., \& Irving, A.~J. 2019, Meteoritics and Planetary Science, 54, 3036, \dodoi{10.1111/maps.13404}

\bibitem[{Garcia {et~al.}(2021)Garcia, Cavanaugh, \& Kacar}]{Garcia2021}
Garcia, A.~K., Cavanaugh, C.~M., \& Kacar, B. 2021, ISME Journal, 15, 2183, \dodoi{10.1038/s41396-021-00971-5}

\bibitem[{Gardner {et~al.}(2006)Gardner, Mather, Clampin, Doyon, Greenhouse, Hammel, Hutchings, Jakobsen, Lilly, Long, Lunine, McCaughrean, Mountain, Nella, Rieke, Rieke, Rix, Smith, Sonneborn, Stiavelli, Stockman, Windhorst, \& Wright}]{Gardner2006}
Gardner, J.~P., Mather, J.~C., Clampin, M., {et~al.} 2006, Space Science Reviews, 123, 485, \dodoi{10.1007/s11214-006-8315-7}

\bibitem[{Glidden {et~al.}(2022)Glidden, Seager, Huang, Petkowski, \& Ranjan}]{Glidden2022}
Glidden, A., Seager, S., Huang, J., Petkowski, J.~J., \& Ranjan, S. 2022, The Astrophysical Journal, 930, 62, \dodoi{10.3847/1538-4357/ac625f}

\bibitem[{Gordon {et~al.}(2017)Gordon, Rothman, Hill, Kochanov, Tan, Bernath, Birk, Boudon, Campargue, Chance, Drouin, Flaud, Gamache, Hodges, Jacquemart, Perevalov, Perrin, Shine, Smith, Tennyson, Toon, Tran, Tyuterev, Barbe, Cs{\'{a}}sz{\'{a}}r, Devi, Furtenbacher, Harrison, Hartmann, Jolly, Johnson, Karman, Kleiner, Kyuberis, Loos, Lyulin, Massie, Mikhailenko, Moazzen-Ahmadi, M{\"{u}}ller, Naumenko, Nikitin, Polyansky, Rey, Rotger, Sharpe, Sung, Starikova, Tashkun, Auwera, Wagner, Wilzewski, Wcis{\l}o, Yu, \& Zak}]{Gordon2017}
Gordon, I.~E., Rothman, L.~S., Hill, C., {et~al.} 2017, Journal of Quantitative Spectroscopy and Radiative Transfer, 203, 3, \dodoi{10.1016/j.jqsrt.2017.06.038}

\bibitem[{Gordon {et~al.}(2022)Gordon, Rothman, Hargreaves, Hashemi, Karlovets, Skinner, Conway, Hill, Kochanov, Tan, Wcis{\l}o, Finenko, Nelson, Bernath, Birk, Boudon, Campargue, Chance, Coustenis, Drouin, Flaud, Gamache, Hodges, Jacquemart, Mlawer, Nikitin, Perevalov, Rotger, Tennyson, Toon, Tran, Tyuterev, Adkins, Baker, Barbe, Can{\`{e}}, Cs{\'{a}}sz{\'{a}}r, Dudaryonok, Egorov, Fleisher, Fleurbaey, Foltynowicz, Furtenbacher, Harrison, Hartmann, Horneman, Huang, Karman, Karns, Kassi, Kleiner, Kofman, Kwabia-Tchana, Lavrentieva, Lee, Long, Lukashevskaya, Lyulin, Makhnev, Matt, Massie, Melosso, Mikhailenko, Mondelain, M{\"{u}}ller, Naumenko, Perrin, Polyansky, Raddaoui, Raston, Reed, Rey, Richard, T{\'{o}}bi{\'{a}}s, Sadiek, Schwenke, Starikova, Sung, Tamassia, Tashkun, Vander~Auwera, Vasilenko, Vigasin, Villanueva, Vispoel, Wagner, Yachmenev, \& Yurchenko}]{Gordon2022}
Gordon, I.~E., Rothman, L.~S., Hargreaves, R.~J., {et~al.} 2022, Journal of Quantitative Spectroscopy and Radiative Transfer, 277, \dodoi{10.1016/j.jqsrt.2021.107949}

\bibitem[{Grant {et~al.}(2023)Grant, van Dishoeck, Tabone, Gasman, Henning, Kamp, G{\"{u}}del, Lagage, Bettoni, Perotti, Christiaens, Samland, Arabhavi, Argyriou, Abergel, Absil, Barrado, Boccaletti, Bouwman, o~Garatti, Geers, Glauser, Guadarrama, Jang, Kanwar, Lahuis, Morales-Calder{\'{o}}n, Mueller, Nehm{\'{e}}, Olofsson, Pantin, Pawellek, Ray, Rodgers-Lee, Scheithauer, Schreiber, Schwarz, Temmink, Vandenbussche, Vlasblom, Waters, Wright, Colina, Greve, Justannont, \& {\"{O}}stlin}]{Grant2023}
Grant, S.~L., van Dishoeck, E.~F., Tabone, B., {et~al.} 2023, The Astrophysical Journal Letters, 947, L6, \dodoi{10.3847/2041-8213/acc44b}

\bibitem[{Hare {et~al.}(2018)Hare, Loftus, Jeffrey, \& Ramsey}]{Hare2018}
Hare, V.~J., Loftus, E., Jeffrey, A., \& Ramsey, C.~B. 2018, Nature Communications, 9, \dodoi{10.1038/s41467-017-02691-x}

\bibitem[{Havig {et~al.}(2011)Havig, Raymond, Meyer-Dombard, Zolotova, \& Shock}]{Havig2011}
Havig, J., Raymond, J., Meyer-Dombard, D., Zolotova, N., \& Shock, E. 2011, Journal of Geophysical Research (Biogeosciences), 116, \dodoi{10.1029/2010JG001415}

\bibitem[{Hayes(2001)}]{Hayes2001}
Hayes, J.~M. 2001, Reviews in Mineralogy and Geochemistry, 43, 225, \dodoi{10.2138/gsrmg.43.1.225}

\bibitem[{House {et~al.}(2022)House, Wong, Webster, Flesch, Franz, Stern, Pavlov, Atreya, Eigenbrode, Gilbert, Hofmann, Millan, Steele, Glavin, Malespin, Mahaffy, \& performed research}]{House2022}
House, C.~H., Wong, G.~M., Webster, C.~R., {et~al.} 2022, Proceedings of the National Academy of Sciences, \dodoi{10.1073/pnas.2115651119/-/DCSupplemental}

\bibitem[{Hu {et~al.}(2015)Hu, Kass, Ehlmann, \& Yung}]{Hu2015}
Hu, R., Kass, D.~M., Ehlmann, B.~L., \& Yung, Y.~L. 2015, Nature Communications, 6, 1, \dodoi{10.1038/ncomms10003}

\bibitem[{Jakosky(2019)}]{Jakosky2019}
Jakosky, B.~M. 2019, Planetary and Space Science, 175, 52, \dodoi{10.1016/j.pss.2019.06.002}

\bibitem[{J{\o}rgensen {et~al.}(2016)J{\o}rgensen, Van Der~Wiel, Coutens, Lykke, M{\"{u}}ller, Van~Dishoeck, Calcutt, Bjerkeli, Bourke, Drozdovskaya, Favre, Fayolle, Garrod, Jacobsen, {\"{O}}berg, Persson, \& Wampfler}]{Jorgensen2016}
J{\o}rgensen, J.~K., Van Der~Wiel, M.~H., Coutens, A., {et~al.} 2016, Astronomy and Astrophysics, 595, \dodoi{10.1051/0004-6361/201628648}

\bibitem[{Keeling(1958)}]{Keeling1958}
Keeling, C.~D. 1958, Geochimica et Cosmochimica Acta, 13, 322, \dodoi{10.1016/0016-7037(58)90033-4}

\bibitem[{Keeling \& Piper(2000)}]{Keeling2000}
Keeling, C.~D., \& Piper, S.~C. 2000, {Interannual Variations of Exchanges of Atmospheric CO2 and 13CO2 with the Terrestrial Biosphere and Oceans from 1978 to 2000} (Scripps Institution of Oceanography, University of California, San Diego, California)

\bibitem[{Kempton {et~al.}(2018)Kempton, Bean, Louie, Deming, Koll, Mansfield, Christiansen, L{\'{o}}pez-Morales, Swain, Zellem, Ballard, Barclay, Barstow, Batalha, Beatty, Berta-Thompson, Birkby, Buchhave, Charbonneau, Cowan, Crossfield, De~Val-Borro, Dragomir, Heng, Hu, Kane, Kreidberg, Mallonn, Morley, Narita, Nascimbeni, Pall{\'{e}}, Quintana, Rauscher, Seager, Shkolnik, Sing, Sozzetti, Stassun, Essen, \& Valenti}]{Kempton2018}
Kempton, E.~M., Bean, J.~L., Louie, D.~R., {et~al.} 2018, Publications of the Astronomical Society of the Pacific, 130, 1, \dodoi{10.1088/1538-3873/aadf6f}

\bibitem[{Kiang {et~al.}(2007)Kiang, Segura, Tinetti, {Govindjee}, Blankenship, Cohen, Siefert, Crisp, \& Meadows}]{Kiang2007}
Kiang, N.~Y., Segura, A., Tinetti, G., {et~al.} 2007, Astrobiology, 7, 252, \dodoi{10.1089/ast.2006.0108}

\bibitem[{Kirkels {et~al.}(2022)Kirkels, De~Boer, Concha~Hern{\'{a}}ndez, Martes, Van Der~Meer, Basu, Usman, \& Peterse}]{Kirkels2022}
Kirkels, F.~M., De~Boer, H.~J., Concha~Hern{\'{a}}ndez, P., {et~al.} 2022, Biogeosciences, 19, 4107, \dodoi{10.5194/bg-19-4107-2022}

\bibitem[{Kobayashi {et~al.}(2011)Kobayashi, Karakas, \& Umeda}]{Kobayashi2011}
Kobayashi, C., Karakas, A.~I., \& Umeda, H. 2011, Monthly Notices of the Royal Astronomical Society, 414, 3231, \dodoi{10.1111/j.1365-2966.2011.18621.x}

\bibitem[{Kochanov {et~al.}(2016)Kochanov, Gordon, Rothman, Wcislo, Hill, \& Wilzewski}]{Kochanov2016}
Kochanov, R.~V., Gordon, I.~E., Rothman, L.~S., {et~al.} 2016, in 71st International Symposium on Molecular Spectroscopy, Vol. 168, 1097--105.
\newblock \url{www.hitran.org}

\bibitem[{Kunimoto {et~al.}(2022)Kunimoto, Winn, Ricker, \& Vanderspek}]{Kunimoto2022Predicting17}
Kunimoto, M., Winn, J., Ricker, G.~R., \& Vanderspek, R.~K. 2022, The Astronomical Journal, 163, 290, \dodoi{10.3847/1538-3881/ac68e3}

\bibitem[{Kurganova \& Lopes~de Gerenyu(2015)}]{Kurganova2015}
Kurganova, I.~N., \& Lopes~de Gerenyu, V.~O. 2015, Eurasian Soil Science, 48, 1009, \dodoi{10.1134/S1064229315090082}

\bibitem[{Langer {et~al.}(1984)Langer, Graedel, Laboratories, Jet, \& Armentrout}]{Langer1984}
Langer, W.~D., Graedel, T.~E., Laboratories, B., Jet, M. A.~F., \& Armentrout, P.~B. 1984, The Astrophysical Journal, 277, 581, \dodoi{10.1086/161730}

\bibitem[{Leman {et~al.}(2004)Leman, Orgel, \& Ghadiri}]{Leman2004}
Leman, L., Orgel, L., \& Ghadiri, M.~R. 2004, Science, 306, 283, \dodoi{10.1126/science.1102722}

\bibitem[{Leshin {et~al.}(2013)Leshin, Mahaffy, Webster, Cabane, Coll, Conrad, Archer, Atreya, Brunner, Buch, Eigenbrode, Flesch, Franz, Freissinet, Glavin, Mcadam, Miller, Ming, Morris, Navarro-Gonz{\'{a}}lez, Niles, Owen, Pepin, Squyres, Steele, Stern, Summons, Sumner, Sutter, Szopa, Teinturier, Trainer, Wray, \& Grotzinger}]{Leshin2013}
Leshin, L.~A., Mahaffy, P.~R., Webster, C.~R., {et~al.} 2013, Science, 341, \dodoi{10.1126/science.12389}

\bibitem[{Lincowski {et~al.}(2019)Lincowski, Lustig-Yaeger, \& Meadows}]{Lincowski2019}
Lincowski, A.~P., Lustig-Yaeger, J., \& Meadows, V.~S. 2019, The Astronomical Journal, 158, 26, \dodoi{10.3847/1538-3881/ab2385}

\bibitem[{Line {et~al.}(2021)Line, Brogi, Bean, Gandhi, Zalesky, Parmentier, Smith, Mace, Mansfield, Kempton, Fortney, Shkolnik, Patience, Rauscher, D{\'{e}}sert, \& Wardenier}]{Line2021}
Line, M.~R., Brogi, M., Bean, J.~L., {et~al.} 2021, Nature, 598, 580, \dodoi{10.1038/s41586-021-03912-6}

\bibitem[{Liuzzi {et~al.}(2022)Liuzzi, Villanueva, Stone, Faggi, Kofman, \& Trompet}]{Liuzzi2022}
Liuzzi, G., Villanueva, G.~L., Stone, S.~W., {et~al.} 2022, in Seventh International Workshop on the Mars Atmosphere: Modelling and Observations, 3532, \dodoi{10.5194/epsc2022-575}

\bibitem[{Lyons {et~al.}(2018)Lyons, Gharib-Nezhad, \& Ayres}]{Lyons2018}
Lyons, J.~R., Gharib-Nezhad, E., \& Ayres, T.~R. 2018, Nature Communications, 9, \dodoi{10.1038/s41467-018-03093-3}

\bibitem[{Madhusudhan(2012)}]{Madhusudhan2012}
Madhusudhan, N. 2012, Astrophysical Journal, 758, \dodoi{10.1088/0004-637X/758/1/36}

\bibitem[{Mahaffy {et~al.}(2013)Mahaffy, Webster, Atreya, Franz, Wong, Conrad, Harpold, Jones, Leshin, Manning, Owen, Pepin, Squyres, \& Trainer}]{Mahaffy2013}
Mahaffy, P.~R., Webster, C.~R., Atreya, S.~K., {et~al.} 2013, Science Team Source: Science, 341, 263, \dodoi{10.1126/science.l237961}

\bibitem[{Mattey(1991)}]{Mattey1991}
Mattey, D.~P. 1991, Geochimica et Cosmochimica Acta, 55, 3467, \dodoi{10.1016/0016-7037(91)90508-3}

\bibitem[{McClure {et~al.}(2023)McClure, Rocha, Pontoppidan, Crouzet, Chu, Dartois, Lamberts, Noble, Pendleton, Perotti, Qasim, Rachid, Smith, Sun, Beck, Boogert, Brown, Caselli, Charnley, Cuppen, Dickinson, Drozdovskaya, Egami, Erkal, Fraser, Garrod, Harsono, Ioppolo, Jim{\'{e}}nez-Serra, Jin, J{\o}rgensen, Kristensen, Lis, McCoustra, McGuire, Melnick, {\"{O}}berg, Palumbo, Shimonishi, Sturm, van Dishoeck, \& Linnartz}]{McClure2023}
McClure, M.~K., Rocha, W.~R., Pontoppidan, K.~M., {et~al.} 2023, Nature Astronomy, 7, 431, \dodoi{10.1038/s41550-022-01875-w}

\bibitem[{McNevin {et~al.}(2007)McNevin, Badger, Whitney, Von~Caemmerer, Tcherkez, \& Farquhar}]{McNevin2007}
McNevin, D.~B., Badger, M.~R., Whitney, S.~M., {et~al.} 2007, Journal of Biological Chemistry, 282, 36068, \dodoi{10.1074/jbc.M706274200}

\bibitem[{Meadows {et~al.}(2018)Meadows, Reinhard, Arney, Parenteau, Schwieterman, Domagal-Goldman, Lincowski, Stapelfeldt, Rauer, Dassarma, Hegde, Narita, Deitrick, Lustig-Yaeger, Lyons, Siegler, \& Grenfell}]{Meadows2018}
Meadows, V.~S., Reinhard, C.~T., Arney, G.~N., {et~al.} 2018, Astrobiology, 18, 630, \dodoi{10.1089/ast.2017.1727}

\bibitem[{Milam {et~al.}(2005)Milam, Savage, Brewster, Ziurys, \& Wyckoff}]{Milam2005}
Milam, S.~N., Savage, C., Brewster, M.~A., Ziurys, L.~M., \& Wyckoff, S. 2005, The Astrophysical Journal, 634, 1126, \dodoi{10.1086/497123}

\bibitem[{Mojzsis {et~al.}(1996)Mojzsis, Arrhenius, McKeegan, Harrison, Nutman, \& Friend}]{Mojzsis1996}
Mojzsis, S.~J., Arrhenius, G., McKeegan, K.~D., {et~al.} 1996, Nature, 384, 55, \dodoi{10.1038/384055a0}

\bibitem[{Molliere \& Snellen(2019)}]{MolliereSnellen2019}
Molliere, P., \& Snellen, I.~A. 2019, Astronomy and Astrophysics, 622, 139, \dodoi{10.1051/0004-6361/201834169}

\bibitem[{Molli{\`{e}}re {et~al.}(2017)Molli{\`{e}}re, van Boekel, Bouwman, Henning, Lagage, \& Min}]{Molliere2017}
Molli{\`{e}}re, P., van Boekel, R., Bouwman, J., {et~al.} 2017, Astronomy {\&} Astrophysics, 600, A10

\bibitem[{Molli{\`{e}}re {et~al.}(2019)Molli{\`{e}}re, Wardenier, Van~Boekel, Henning, Molaverdikhani, \& Snellen}]{Molliere2019prt}
Molli{\`{e}}re, P., Wardenier, J.~P., Van~Boekel, R., {et~al.} 2019, Astronomy and Astrophysics, 627, \dodoi{10.1051/0004-6361/201935470}

\bibitem[{Mooshammer {et~al.}(2020)Mooshammer, Alves, Bayer, Melcher, Stieglmeier, Jochum, Rittmann, Watzka, Schleper, Herndl, \& Wanek}]{Mooshammer2020}
Mooshammer, M., Alves, R.~J., Bayer, B., {et~al.} 2020, Frontiers in Microbiology, 11, \dodoi{10.3389/fmicb.2020.01710}

\bibitem[{Moreras-Marti {et~al.}(2022)Moreras-Marti, Fox-Powell, Cousins, Macey, \& Zerkle}]{Moreras-Marti2022}
Moreras-Marti, A., Fox-Powell, M., Cousins, C.~R., Macey, M.~C., \& Zerkle, A.~L. 2022, Journal of the Geological Society, 179, \dodoi{10.1144/jgs2021-134}

\bibitem[{Morley {et~al.}(2019)Morley, Skemer, Miles, Line, Lopez, Brogi, Freedman, \& Marley}]{Morley2019}
Morley, C.~V., Skemer, A.~J., Miles, B.~E., {et~al.} 2019, The Astrophysical Journal, 882, L29, \dodoi{10.3847/2041-8213/ab3c65}

\bibitem[{Nealson \& Rye(2003)}]{Nealson2003}
Nealson, K.~H., \& Rye, R. 2003, in Treatise on Geochemistry, ed. W.~H. Schlesinger, H.~D. Holland, \& K.~K. Turekian, Vol.~8, 682, \dodoi{10.1016/B0-08-043751-6/08126-3}

\bibitem[{Neveu {et~al.}(2018)Neveu, Hays, Voytek, New, \& Schulte}]{Neveu2018}
Neveu, M., Hays, L.~E., Voytek, M.~A., New, M.~H., \& Schulte, M.~D. 2018, Astrobiology, 18, 1375, \dodoi{10.1089/ast.2017.1773}

\bibitem[{Niles {et~al.}(2010)Niles, Boynton, Hoffman, Ming, \& Hamara}]{Niles2010}
Niles, P.~B., Boynton, W.~V., Hoffman, J.~H., Ming, D.~W., \& Hamara, D. 2010, {Stable Isotope Measurements of Martian Atmospheric CO2 at the Phoenix Landing Site}, Tech. Rep. 5997

\bibitem[{{NOAA}(2020)}]{NOAAGlobalMonitoringLab}
{NOAA}. 2020, {NOAA Global Monitoring Lab: Earth Systems Research Laboratories}.
\newblock \url{https://www.esrl.noaa.gov/gmd/dv/iadv/}

\bibitem[{Nomura {et~al.}(2022)Nomura, Furuya, Cordiner, Charnley, Alexander, Nixon, Guzman, Yurimoto, Tsukagoshi, \& Iino}]{Nomura2022TheSystem}
Nomura, H., Furuya, K., Cordiner, M.~A., {et~al.} 2022, \dodoi{10.48550/arXiv.2203.10863}

\bibitem[{O'Leary(1988)}]{OLeary1988}
O'Leary, M.~H. 1988, BioScience, 38, 328, \dodoi{10.2307/1310735}

\bibitem[{Olson {et~al.}(2018)Olson, Schwieterman, Reinhard, Ridgwell, Kane, Meadows, \& Lyons}]{Olson2018}
Olson, S.~L., Schwieterman, E.~W., Reinhard, C.~T., {et~al.} 2018, The Astrophysical Journal, 858, L14, \dodoi{10.3847/2041-8213/aac171}

\bibitem[{Owen(1982)}]{Owen1982}
Owen, T. 1982, Adv. Space Res, 2, 75, \dodoi{10.1016/0273-1177(82)90107-7}

\bibitem[{Petkowski {et~al.}(2020)Petkowski, Bains, \& Seager}]{Petkowski2020}
Petkowski, J.~J., Bains, W., \& Seager, S. 2020, Life, 10, 1, \dodoi{10.3390/life10060084}

\bibitem[{Ricci {et~al.}(2023)Ricci, Fiebig, Tassi, Hofmann, Capecchiacci, \& Vaselli}]{Ricci2023}
Ricci, A., Fiebig, J., Tassi, F., {et~al.} 2023, Geochimica et Cosmochimica Acta, \dodoi{https://doi.org/10.1016/j.gca.2023.10.019}

\bibitem[{Ricker {et~al.}(2014)Ricker, Winn, Vanderspek, Latham, Bakos, Bean, Berta-Thompson, Brown, Buchhave, Butler, Butler, Chaplin, Charbonneau, Christensen-Dalsgaard, Clampin, Deming, Doty, De~Lee, Dressing, Dunham, Endl, Fressin, Ge, Henning, Holman, Howard, Ida, Jenkins, Jernigan, Johnson, Kaltenegger, Kawai, Kjeldsen, Laughlin, Levine, Lin, Lissauer, MacQueen, Marcy, McCullough, Morton, Narita, Paegert, Palle, Pepe, Pepper, Quirrenbach, Rinehart, Sasselov, Sato, Seager, Sozzetti, Stassun, Sullivan, Szentgyorgyi, Torres, Udry, \& Villasenor}]{Ricker2014}
Ricker, G.~R., Winn, J.~N., Vanderspek, R., {et~al.} 2014, Journal of Astronomical Telescopes, Instruments, and Systems, 1, 014003, \dodoi{10.1117/1.jatis.1.1.014003}

\bibitem[{Rodler \& L{\'{o}}pez-Morales(2014)}]{Rodler2014}
Rodler, F., \& L{\'{o}}pez-Morales, M. 2014, Astrophysical Journal, 781, \dodoi{10.1088/0004-637X/781/1/54}

\bibitem[{Rothman {et~al.}(1998)Rothman, Rinsland, Goldman, Massie, Edwards, Flaud, Perrin, Camy-Peyret, Dana, Mandin, Schroeder, McCann, Gamache, Wattson, Yoshino, Chance, Jucks, Brown, Nemtchinov, \& Varanasi}]{Rothman1996}
Rothman, L.~S., Rinsland, C.~P., Goldman, A., {et~al.} 1998, Journal of Quantitative Spectroscopy and Radiative Transfer, 60, 665, \dodoi{https://doi.org/10.1016/S0022-4073(98)00078-8}

\bibitem[{Rothschild \& Desmarais(1989)}]{Rothschild1989}
Rothschild, L.~J., \& Desmarais, D. 1989, Ads'. Space Ret, 9, 159, \dodoi{10.1016/0273-1177(89)90223-8}

\bibitem[{Rustamkulov {et~al.}(2023)Rustamkulov, Sing, Mukherjee, May, Kirk, Schlawin, Line, Piaulet, Carter, Batalha, Goyal, L{\'{o}}pez-Morales, Lothringer, MacDonald, Moran, Stevenson, Wakeford, Espinoza, Bean, Batalha, Benneke, Berta-Thompson, Crossfield, Gao, Kreidberg, Powell, Cubillos, Gibson, Leconte, Molaverdikhani, Nikolov, Parmentier, Roy, Taylor, Turner, Wheatley, Aggarwal, Ahrer, Alam, Alderson, Allen, Banerjee, Barat, Barrado, Barstow, Bell, Blecic, Brande, Casewell, Changeat, Chubb, Crouzet, Daylan, Decin, D{\'{e}}sert, Mikal-Evans, Feinstein, Flagg, Fortney, Harrington, Heng, Hong, Hu, Iro, Kataria, Kempton, Krick, Lendl, Lillo-Box, Louca, Lustig-Yaeger, Mancini, Mansfield, Mayne, Miguel, Morello, Ohno, Palle, Petit dit de~la Roche, Rackham, Radica, Ramos-Rosado, Redfield, Rogers, Shkolnik, Southworth, Teske, Tremblin, Tucker, Venot, Waalkes, Welbanks, Zhang, \& Zieba}]{Rustamkulov2023}
Rustamkulov, Z., Sing, D.~K., Mukherjee, S., {et~al.} 2023, Nature, 614, 659, \dodoi{10.1038/s41586-022-05677-y}

\bibitem[{Schmidt {et~al.}(2013)Schmidt, Johnson, \& Schinke}]{Schmidt2013}
Schmidt, J.~A., Johnson, M.~S., \& Schinke, R. 2013, Proceedings of the National Academy of Sciences of the United States of America, 110, 17691, \dodoi{10.1073/pnas.1213083110}

\bibitem[{Schrey {et~al.}(1986)Schrey, Rothermel, Kaufl, \& Drapatz}]{Schrey1986}
Schrey, U., Rothermel, H., Kaufl, H., \& Drapatz, S. 1986, A{\&}A, 200

\bibitem[{Seager {et~al.}(2021)Seager, Petkowski, G{\"{u}}nther, Bains, Mikal-evans, \& Deming}]{Seager2021}
Seager, S., Petkowski, J.~J., G{\"{u}}nther, M.~N., {et~al.} 2021, Universe, 7, 172, \dodoi{10.3390/universe7060172}

\bibitem[{Seager {et~al.}(2005)Seager, Turner, Schafer, \& Ford}]{Seager2005}
Seager, S., Turner, E.~L., Schafer, J., \& Ford, E.~B. 2005, Astrobiology, 5, 372, \dodoi{10.1089/ast.2005.5.372}

\bibitem[{Serindag \& Snellen(2019)}]{Serindag2019}
Serindag, D.~B., \& Snellen, I. A.~G. 2019, The Astrophysical Journal, 871, L7, \dodoi{10.3847/2041-8213/aafa1f}

\bibitem[{Sheffer {et~al.}(1992)Sheffer, Federman, Lambert, \& Cardelli}]{Sheffer1992}
Sheffer, Y., Federman, S., Lambert, D.~L., \& Cardelli, J.~A. 1992, The Astrophysical Journal, 397, 482, \dodoi{10.1086/171805}

\bibitem[{Sim {et~al.}(2011)Sim, Bosak, \& Ono}]{Sim2011}
Sim, M.~S., Bosak, T., \& Ono, S. 2011, New Series, 333, 74, \dodoi{10.1126/science.l204394}

\bibitem[{Smith \& Adams(1980)}]{SmithAdams1980}
Smith, D., \& Adams, N.~G. 1980, The Astrophysical Journal, 242, 424

\bibitem[{Smith {et~al.}(2015)Smith, Pontoppidan, Young, \& Morris}]{Smith2015}
Smith, R.~L., Pontoppidan, K.~M., Young, E.~D., \& Morris, M.~R. 2015, Astrophysical Journal, 813, 120, \dodoi{10.1088/0004-637X/813/2/120}

\bibitem[{Snellen {et~al.}(2013)Snellen, De~Kok, Le~Poole, Brogi, \& Birkby}]{Snellen2013}
Snellen, I.~A., De~Kok, R.~J., Le~Poole, R., Brogi, M., \& Birkby, J. 2013, Astrophysical Journal, 764, \dodoi{10.1088/0004-637X/764/2/182}

\bibitem[{Still \& Rastogi(2017)}]{Still2017}
Still, C., \& Rastogi, B. 2017, Journal of Geophysical Research: Biogeosciences, 122, 3108, \dodoi{10.1002/2017JG004155}

\bibitem[{Unkovich(2013)}]{Unkovich2013}
Unkovich, M. 2013, New Phytologist, 198, 643, \dodoi{10.1111/nph.12227}

\bibitem[{Watson {et~al.}(1976)Watson, Anicich, \& Huntress}]{Watson1976}
Watson, W.~D., Anicich, V.~G., \& Huntress, W.~T. 1976, The Astrophysical Journal, 205, 165

\bibitem[{Webster {et~al.}(2013)Webster, Mahaffy, Webster, Flesch, Niles, Jones, Leshin, Atreya, Stern, Christensen, Owen, Franz, Pepin, Steele, \& {MSL Science Team}}]{Webster2013}
Webster, C., Mahaffy, P.~R., Webster, C., {et~al.} 2013, Science, 341, 260, \dodoi{10.1126/science.123796}

\bibitem[{Wieloch {et~al.}(2022)Wieloch, Grabner, Augusti, Serk, Ehlers, Yu, \& Schleucher}]{Wieloch2022}
Wieloch, T., Grabner, M., Augusti, A., {et~al.} 2022, New Phytologist, 234, 449, \dodoi{10.1111/nph.18014}

\bibitem[{Woods(2009)}]{Woods2009}
Woods, P.~M. 2009, {Carbon Isotope Measurements in the Solar System}, Tech. rep., \dodoi{10.48550/arXiv.0901.4513}

\bibitem[{Woods \& Willacy(2009)}]{WoodsWillacy2009}
Woods, P.~M., \& Willacy, K. 2009, Astrophysical Journal, 693, 1360, \dodoi{10.1088/0004-637X/693/2/1360}

\bibitem[{Yoshida {et~al.}(2023)Yoshida, Aoki, Ueno, Terada, Nakamura, Shiobara, Yoshida, Nakagawa, Sakai, \& Koyama}]{Yoshida2023}
Yoshida, T., Aoki, S., Ueno, Y., {et~al.} 2023, The Planetary Science Journal, 4, 53, \dodoi{10.3847/PSJ/acc030}

\bibitem[{Yoshida {et~al.}(2022)Yoshida, Nomura, Furuya, Tsukagoshi, \& Lee}]{Yoshida2022}
Yoshida, T.~C., Nomura, H., Furuya, K., Tsukagoshi, T., \& Lee, S. 2022, The Astrophysical Journal, 932, 126, \dodoi{10.3847/1538-4357/ac6efb}

\bibitem[{Yurimoto \& Kuramoto(2004)}]{YurimotoKuramoto2004}
Yurimoto, H., \& Kuramoto, K. 2004, Science, 305, 1763, \dodoi{10.1126/science.1100989}

\bibitem[{Zhan {et~al.}(2021)Zhan, Seager, Petkowski, Sousa-Silva, Ranjan, Huang, \& Bains}]{Zhan2021}
Zhan, Z., Seager, S., Petkowski, J.~J., {et~al.} 2021, Astrobiology, 21, \dodoi{10.1089/ast.2019.2146}

\bibitem[{Zhang {et~al.}(2021)Zhang, Snellen, Bohn, Molli{\`{e}}re, Ginski, Hoeijmakers, Kenworthy, Mamajek, Meshkat, Reggiani, \& Snik}]{Zhang2021}
Zhang, Y., Snellen, I.~A., Bohn, A.~J., {et~al.} 2021, Nature, 595, 370, \dodoi{10.1038/s41586-021-03616-x}

\bibitem[{Zyakun {et~al.}(2009)Zyakun, Lunina, Prusakova, Pimenov, \& Ivanov}]{Zyakun2009}
Zyakun, A.~M., Lunina, O.~N., Prusakova, T.~S., Pimenov, N.~V., \& Ivanov, M.~V. 2009, Microbiology, 78, 757, \dodoi{10.1134/S0026261709060137}

\end{thebibliography}
\bibliographystyle{aasjournal}

\end{document}